\documentclass[pra,twocolumn,floatfix]{revtex4}
\usepackage{graphicx}
\usepackage{amssymb}
\usepackage{tikz}
\usepackage{pgfplots}
\usepackage{epstopdf} 
\usetikzlibrary{pgfplots.groupplots}

\newcommand{\be}{\begin{eqnarray}}
\newcommand{\ee}{\end{eqnarray}}

\newcommand{\vA}{{\bf A}}

\newcommand{\vE}{{\bf E}}
\newcommand{\vf}{{\bf f}}
\newcommand{\Vr}{{\bf r}}
\newcommand{\vx}{{\bf x}}

\newcommand{\ffp}{p}

\newcommand{\lpd}[2]{\partial #1 / \partial #2}

\newcommand{\dby}[2]{ \frac{{\rm d} #1}{{\rm d} #2}}
\newcommand{\dtau}{\rm d\tau}

\newcommand{\dS}{{\rm d}S}
\newcommand{\Fto}{{_2  {\rm F}_{\! 1}}}

\newcommand{\ppx}[1]{\frac{\partial #1}{\partial x}}
\newcommand{\binomial}[2]{\left( \!\! \begin{array}{c} {#1}\\{#2} \end{array} \!\! \right) }

\newcommand{\Grad}{\mbox{\boldmath $\nabla$}}

\newcommand{\Lap}{\nabla^2}

\newcommand{\myfig}[2] {\centerline{\resizebox{!}{#1\textwidth}{\includegraphics{#2}}} }

\def\localfiles{1}

\newcommand{\boxtable}[3]{\begin{table*}[t!]
\fbox{\hspace{0.5mm}
\begin{minipage}[t]{0.96\textwidth} {#2} \end{minipage}
\hspace{1mm}}
\caption{#3}
\label{#1}
\end{table*}
}

\begin{document}

\title{Electromagnetic self-force for axially symmetric charge on a spherical shell}

\author{Andrew M. Steane}
%\ead{a.steane@physics.ox.ac.uk}
%\affiliation{Department of Atomic and Laser Physics, Clarendon Laboratory, Parks Road, Oxford OX1 3PU, England.}
\address{Department of Atomic and Laser Physics, Clarendon Laboratory, Parks Road, Oxford OX1 3PU, England.}
\date{\today}

\label{firstpage}

\begin{abstract}
%{self-force, radiation reaction, hyperbolic motion, multipole, Rindler frame}
We obtain the fields and electromagnetic self-force of a charge distributed on the surface of a sphere
undergoing rigid motion at constant proper acceleration, where the charge distribution has axial symmetry about
the direction of motion. A closed-form expression for the self-force is given in
terms of the multipole moments of the charge distribution. Applications to the
electrodynamics of a dipole, and to electromagnetic self-force near 
a horizon (in spacetime) are discussed.
\end{abstract}

\maketitle

\section{Introduction}

This work has two main themes: the understanding of self-force, and the discovery of exact
analytical results in classical electromagnetism.

The study of self-force and radiation reaction in classical electromagnetism has a long history
(\cite{05Abraham,15Schott,38Dirac,90Rohrlich,97Rohrlich,98Jackson,04Spohn,14SteaneB,91Ford,93Ford,12OConnell}). 
The subject is important to learning how concepts as basic as mass, momentum and energy are
to be understood when fields and continuous media are involved and an exact relativistic formulation
is sought. This in turn feeds into the understanding of renormalization and radiation reaction
in quantum physics and general relativity. Also, high-precision mass measurements are now
reaching the precision where the electromagnetic contribution to the mass of atoms and molecules
can in principle be detected \cite{04Rainville}. Electromagnetic self-force
is also of practical relevance in experiments involving high acceleration in the fields of high-power
laser pulses \cite{14Burton}.

Concerning exact analytical results, Heaviside remarked, in connection with the calculation of
fields in electromagnetism (and one could extend the remark to theoretical physics more generally),
``for it is very exceptional to arrive at simple results." 
There are only a few cases where the electromagnetic field of a physical system can be calculated exactly,
and these mostly involve only inertial motion. When charged bodies accelerate, their associated field cannot be calculated
until the motion is specified sufficiently fully. When self-force is non-negligible this motion itself depends
on the fields one wishes to obtain, so that typically one has a non-linear problem and there is no route
to a full analytical solution in closed form. 
Indeed, no problem in classical electromagnetism
has ever been analysed in full, when accelerated motion is involved, because the equations are too 
difficult to solve. In this sense, within the assumptions of classical physics,
we know the differential equations that describe how charged things behave, 
but we do not know (exactly) how {\em any} charged thing behaves, unless it is moving inertially! 

For a long time an equation of motion for
a point-like charged body was sought (that is to say, an equation which took self-force into account
and would be valid in the limit $R \rightarrow 0$ where $R$ is the size of the body). This quest turned
on a misunderstanding, however. One cannot treat the accelerating body as point-like, because this is 
unphysical if the mass and charge are finite \cite{14SteaneB,09Gralla,48Bohm,61Erber,91Ford}. 
Any exact calculation of self-force, for a body of non-vanishing charge, must therefore treat 
a body of non-infinitesimal size. One is then dealing with a world-tube as opposed to a world-line,
and in most cases the calculation is not tractable.

In order to solve the problem of motion under a force in general, one would specify the applied force,
and the equation for the internal dynamics of the body in question (an equation of state), and set
out a set of integrals such as those presented by Harte \cite{15Harte}. The sequence of calculations
involves both finding the world-tube of the body, and hence its fields, and feeding this information
back into the integrals. It has not proved possible yet to find an exact solution in closed form when
the problem is set out this way.

In order to simplify the calculation, one may adopt the strategy of specifying the world-tube at the
outset. One simply assumes that a charged body is moving in some specified way, and calculates
the self-force that results. A suitable strategy, for example, is to assume that
the body in question is moving rigidly \cite{64Nodvik,14Lyle,12Steane}. 
By {\em rigid} motion here we mean that the response of
the body to the forces on it is such that its size and shape is constant. That is to say,
one may pick a worldline to serve as reference, and at each moment of proper time $\tau$ evaluated along the
reference worldline, there exists an inertial reference frame $S(\tau)$ 
in which all parts of the body are at rest, and the distance between any two parts of the body, evaluated
in $S(\tau)$, is independent of $\tau$. As Nodvik puts it, ``the non-electromagnetic forces necessary for stability
are taken into account implicitly by requiring that the charge distribution retain a given shape throughout the course
of its motion." By assuming rigid motion so defined, one is assuming a somewhat artificial
situation, in which the net result of the initial conditions and the internal and external forces is this kind of motion,
but this is a physically possible case and it is, arguably, the simplest type of accelerated motion. It merits
consideration as a canonical case that we do well to understand, if we can. 

Rigid motion is also interesting because rigid motion at constant proper acceleration
corresponds to a static situation in the {\em Rindler frame}
of general relativity \cite{06Rindler,12Steane}. 
This frame plays an important role in the consideration of such concepts as horizons,
the equivalence principle, Unruh radiation and vacuum entanglement.

Nodvik presented a tour-de-force calculation which treated rigid motion of a spherically
symmetric charge distribution undergoing otherwise arbitrary motion. \cite{64Nodvik}
The proper acceleration, for example,
was not assumed to be constant.  He obtained the lowest four terms in an expansion of the self-force
in integrals related to radial moments of the charge distribution. Previous authors had obtained
only the lowest two terms.
In the case of a charged sphere of radius $R$ and proper acceleration $g$, the series expansion 
converges rapidly when $g R \ll c^2$.

The present work provides exact results in the case of rigid motion at constant proper acceleration,
for a charge distribution which is confined to the surface of a shell of given radius, and has axial
symmetry about the line along which the shell is accelerating. Thus we have a more restrictive
assumption about the motion than the one adopted by Nodvik, but we do not require
spherical symmetry of the charge distribution, and we obtain exact results---that is,
expressions that include all orders in an expansion parameter. 
This builds on studies described in \cite{14Steane,Steane15}. 

In \cite{14Steane} and \cite{Steane15} the 
electromagnetic self-force was calculated exactly for the case of
a uniform spherical shell undergoing rigid motion at constant proper acceleration. 
Until now, this is the only case for which an exact result is known for self-force in electromagnetism 
(except the trivial case of zero for a body moving inertially).  
To be precise, the analysis in \cite{14Steane,Steane15} furnishes a power series whose sum is the self-force.
An explicit expression for a general coefficient in the series is furnished, and the sum
can be obtained in terms of the inverse hyperbolic tangent and Lerch transcendent functions
(see the appendix to this paper). This expression has been checked to high 
order in the expansion parameter $(gR/c^2)$, and it is reasonable to conjecture that it is valid at all orders, 
but the method of calculation did not furnish a proof of this. If one accepts the conjecture then 
one may say that up till now the electromagnetic self-force, in classical electromagnetism,
has been obtained for one physical system: the constantly accelerating rigid spherical shell
with uniformly distributed charge. 
The present work generalizes this to such a shell with an axially symmetric but otherwise
arbitrary distribution of charge. Among other things, this enables one to treat, exactly, a
shell with a non-zero electric dipole moment. 
This is a case having further interest because 
it is associated with two paradoxes outlined in section \ref{s.dipole}.

The result of the present type of study is 
an explicit formula for one of the significant items (the self-force owing to the field sourced by the body's
electric charge), but it does not on its own solve the dynamics in
full, because it leaves unspecified the stress in the body which must be present in order for the shell
to undergo the assumed motion. However, a further useful property was shown in \cite{Steane15}, namely that
when the net force on the extended body is suitably defined, the internal pressure (or tension) makes zero 
contribution to the total force after it has been summed over the surface of the body. Only
the sheer stress contributes.

The text is laid out as follows. Sections \ref{s.term}, \ref{s.outline} introduce
terminology and outline the method of calculation. Section \ref{s.field} presents the
calculation of the electric potential, and hence the field, throughout the interior of
the spherical shell, through the use of an expansion in a set of basis potentials
which satisfy a suitable differential equation.
Section \ref{s.self} uses this to obtain the self-force in the case
of a charge distribution having only one non-zero multipole moment. The
self-force is found by summing an infinite series of contributions, each of which can be
obtained exactly, as can the coefficients of the terms in the series. Section
\ref{s.checks} compares these analytical results to some examples of numerical integration,
and to a lowest-order result obtained by Nodvik. Section \ref{s.general} then presents
the generalization to any charge distribution having axial symmetry. Section 
\ref{s.application} comments on applications to study of
the self-force of a dipole and the dependence of self-force on the charge distribution.
The paper concludes with some brief pointers towards the treatment of a completely general
charge distribution.

\section{Terminology}  \label{s.term}

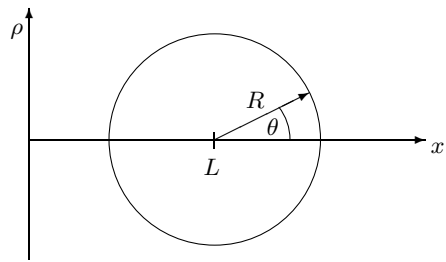
\begin{figure}
\begin{center}
\begin{picture}(200,100)
\put(0,50){\vector(1,0){150} \put(2,-5){$x$}}
\put(0,5){\vector(0,1){95}  \put(-7,85){$\rho$}}
\put(70,50){\vector(2,1){36}}
\put(70,47){\line(0,1){6} \put(-4,-10){$L$}}
\put(82,62){$R$}
\put(90,52){$\theta$}
%\put(70,50){\circle{100}}
\put(27,10){
\begin{tikzpicture}
\draw (1,0) arc (0:38:20 pt);
\draw (0,0) circle (40 pt);
\end{tikzpicture}
}
\end{picture}
\end{center}
\caption{A spherical shell accelerates in the positive $x$-direction, coming momentarily to
rest as shown.}
\label{f.shell}
\end{figure}

We treat a spherical shell of proper radius $R$, undergoing rigid motion (as defined above) with constant
proper acceleration (also known as hyperbolic motion) 
along the $x$ axis. We calculate the fields and force in the instantaneous rest frame.
In this frame, the centre of the shell comes momentarily to rest at $x=L$ and the proper acceleration
of the centre of the shell is $g = c^2/L$. See figure \ref{f.shell}.

The surface charge density is expressed
\be
\sigma(\theta) = \sum_{l=0}^\infty s_l \sigma_l(\theta)         \label{sigseries}
\ee
where $\theta$ is the polar angle in a system of spherical polar coordinates centred at the sphere's centre,
with axis along the direction of constant proper acceleration, 
$s_l$ are constants which describe the distribution, and
\be
\sigma_l = \sigma_0 P_l( \cos \theta )          \label{sig}
\ee
where $\sigma_0$ is a constant and $P_l$ is a Legendre polynomial. Thus we assume an axially
symmetric but otherwise arbitrary distribution of charge on the surface of the shell.

\begin{figure}
\myfig{0.38}{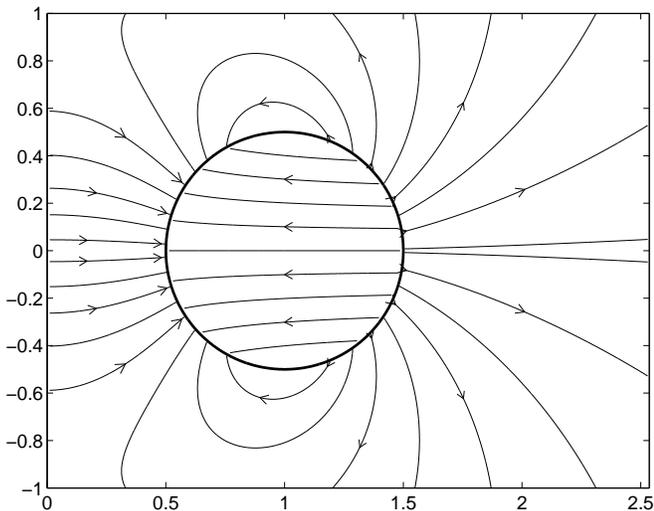}
\caption{The electric field of a spherical shell with charge density $\sigma = \sigma_0 \cos \theta$
undergoing rigid hyperbolic motion. Such a shell has non-zero dipole moment and zero total charge. The field is shown in the instantaneous rest frame, for the case of a sphere with radius $R = (1/2) c^2 / g$ where $g$ is
the proper acceleration of the centre of the sphere.}
\label{f.dipole} 
\end{figure}

\begin{figure}
\myfig{0.35}{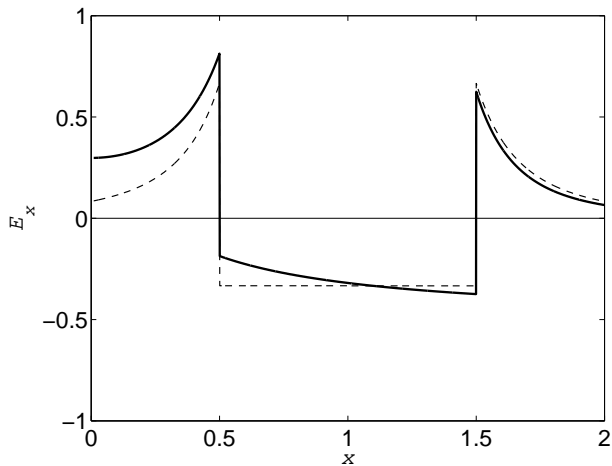}
\caption{The electric field of the sphere shown in figure \ref{f.dipole}, for points on the $x$ axis.
The dashed line shows the result for a non-accelerating sphere, for comparison.}
\label{f.Edip}
\end{figure}

\begin{figure}
\myfig{0.35}{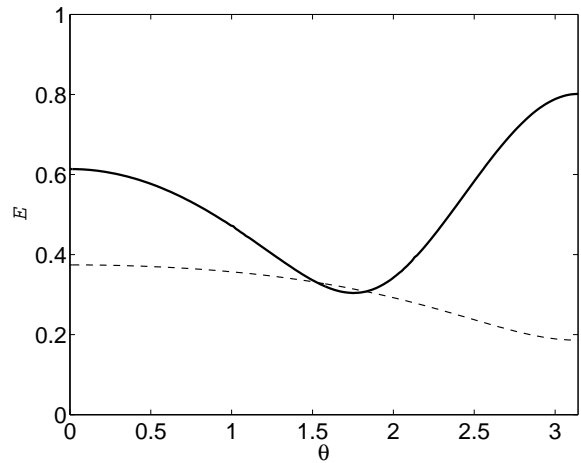}
\caption{The magnitude of the electric field of the sphere shown in figure \ref{f.dipole}, for points on the outside surface
(full line) and inside surface (dashed line) of the shell.}
\label{f.Ediptheta}
\end{figure}

If only one term in the series (\ref{sigseries}) is non-zero then we shall describe the charged object
as a `multipolar sphere'. If such a multipolar sphere were at rest and not accelerating, then 
the electric scalar potential outside it would be
$\phi(r, \theta) = \alpha r^{-(l+1)} P_l( \cos \theta )$ where $\alpha = R^{l+2} \sigma_0 / (2 l + 1) \epsilon_0$,
in the gauge where the vector potential is zero.
In other words, the exterior field of the multipolar sphere is that of a multipole moment of order $l$ (c.f. figures
\ref{f.dipole}--\ref{f.Ediptheta}).

%Even something as simple as the acceleration of a uniformly 
%charged spherical body in a uniform electric field 
%is not known exactly, in that no explicit formula for it has ever been obtained. For
%example, if we specify the conditions to be such that the body has been moving in the field for
%sufficiently long that damping forces
%provided by its own material have reduced any internal oscillations to zero, then the relationship between the
%mass, internal stress, charge, acceleration and applied field has never been found exactly. 

%A significant step towards solving this problem was taken by Steane, who found the electromagnetic 
%self-force for a uniformly charged spherical shell undergoing rigid motion at constant proper
%acceleration. 

%Two strategies have proved useful when one seeks for greater accuracy than is available from a few terms in
%a perturbative expansion. First, one may adopt numerical techniques to solve
%the system of equations to within numerical accuracy. This is sufficient to serve the purely 
%pragmatic aim of prediction and comparison with experiment, and also by numerical exploration one can
%get insight into and intuition about the physical effects.

\section{Outline of the method} \label{s.outline}

We here outline the method that is discussed more fully in 
\cite{14Steane} and \cite{Steane15}. We wish to calculate, first of all, the electric
field, in the instantaneous rest frame, of a body undergoing rigid motion at constant proper acceleration. 
Choose the coordinate axes so that the $x$ axis is parallel to the motion, and $x=0$ is the plane at which the
acceleration tends to infinity. It is shown
in \cite{14Steane}  that, at the moment in question (i.e. at $t=0$ in the inertial frame whose clock is set to zero when the body
is momentarily at rest in the frame), the vector potential satisfies $\lpd{\vA}{t} = (\phi/x, 0, 0)$ where
$\phi$ is the scalar potential, and therefore the electric field can be obtained from $\vE = -\Grad \phi - (\phi/x) \hat{\vx}$.
It is further shown that $\phi$ satisfies
\be
x^2 \Lap \phi + x \ppx{\phi} - \phi = 0.  \label{diffeq}
\ee
The method consists of first finding an infinite set of solutions to this equation, and then expressing the scalar potential
for the specific physical situation as a linear combination of these solutions. The coefficients in the sum are obtained
by calculating $\phi$ along a line (the $x$ axis) by integration over the body, and matching the sum to this.
Here, the value of $\phi(x,0,0)$ is acting as a boundary condition for the solution of the differential equation.
A line rather than a closed surface is sufficient in the case of axial symmetry, because then the
problem reduces to two dimensions. Finally, the self-force is obtained from an integral expressing the force on
the charge distribution owing to the electric field so calculated.

The solutions of (\ref{diffeq}) to be employed are
\be
\phi_n(x,\rho) &=& 
 \sum_{m=0}^n \frac{ (-1)^m }{m+1} \binomial{n}{m}^{\! 2} x^{2m+1} \rho^{2(n-m)}  \nonumber\\
&=& x \rho^{2n} \Fto \! \left(-n,-n,2,-x^2/\rho^2 \right)  \label{Vnxrho}
\ee
where $\rho$ is the radial coordinate in a cylindrical system of coordinates with axis $x$,
$(\stackrel{n}{\mbox{\scriptsize $m$}})=n!/m!(n-m)!$ is a binomial coefficient and 
$\Fto$ is the hypergeometric function. 
The electric field given by these solutions is
\be
E_{x,n} &=&
 -2 \rho^{2n} \Fto\! \left(-n,-n,1,-x^2/\rho^2 \right),        \label{Efieldn}    \\
E_{\rho,n} &=&
-2 n x \rho^{2n-1} \Fto\! \left(1-n,-n,2,-x^2/\rho^2 \right).    \label{Efieldrhon}
\ee

The importance of the above method is two-fold. First, equation (\ref{diffeq}) plays the
role, for the fields we wish to calculate, that Laplace's equation serves in electrostratics, and the solutions
(\ref{Vnxrho}) play the role equivalent to that of the spherical harmonics in treatments of
Laplace's equation. That is to say, this approach yields a powerful and insightful general approach
to studying the fields of a body undergoing rigid hyperbolic motion, and consequently it yields
a powerful general method for treating electromagnetism in the Rindler frame. 

The second importance of the method is that it is tractable. For the case under study, it leads to
tractable integrals, and thus simplifies the calculation in a significant way. The `brute force'
method to find self-force is to write down the integrals describing the force on
each element of charge owing to the field sourced by each other element of charge. Such integrals
are set out in section \ref{s.checks}; we do not need them here except as a way to use 
numerical integration to provide a consistency check on our results.

\section{Interior field}  \label{s.field}

The scalar potential per unit charge, owing to a point charge undergoing hyperbolic motion along the $x$ axis
with proper acceleration $c^2/L$, at the moment when it comes to rest at $x=L$, is given by \cite{60Fulton,00EriksenI,12Steane}
\be
\tilde{\phi}(L;\, x,\,y,\,z) = \frac{ (4 \pi \epsilon_0)^{-1} \left( L^2 + x^2 + y^2+z^2 \right) }
{x [(L^2 + x^2 + y^2+z^2)^2 - 4 L^2 x^2 ]^{1/2}},
\ee
where we have adopted SI units.

%In the following we shall omit the factor $4 \pi \epsilon_0$ without comment, and then reinsert it at the end.

The scalar potential owing to a distribution of charge, moving rigidly with constant proper acceleration
in the $x$ direction, is given by
\begin{widetext}
\be
\phi(x,y,z) = \int\!\! \int\!\! \int {\rm d}x_s {\rm d}y_s {\rm d} z_s
\varrho(x_s,y_s,z_s)
\tilde{\phi}(x_s; \, x,\, y-y_s, \, z-z_s)        \label{phifromtilde}
\ee
at the moment when it comes to rest, where $\varrho$ is the charge density.
The form of this result, especially the dependence on $x$ and $x_s$, is explained in \cite{14Steane}. Hence,
in the case of a spherical shell with surface charge density $\sigma_l(\theta)$,
the scalar potential on the $x$ axis is
\be
\phi(l; \,x,0,0) =
%\frac{1}{2 R} \int_{L-R}^{L+R} \sigma_l( \theta )
%\frac{ R^2 + 2 x_s + x^2 -1}{x \sqrt{(R^2 + 2x_s + x^2-1)^2 - 4 x_s^2 x^2}} {\rm d}x_s
\frac{R}{2 \epsilon_0} \int_{L-R}^{L+R} \sigma_l( \theta )
\frac{ R^2 + 2 L x_s + x^2 -L^2}{x \sqrt{(R^2 + 2 L x_s + x^2-L^2)^2 - 4 x_s^2 x^2}} {\rm d}x_s
\label{phix0}
\ee
where $\theta$ and $L$ are as defined in figure \ref{f.shell}, so that $x_s = L + R \cos \theta$.

Evaluating the integral, one finds
\be
\phi(l;\,x,0,0) = 
%\frac{ (s + R^2) \sigma_0 / \epsilon_0}{2^{3l+2} x s^{l+1} R^{l-1}} \!
%\left[  p_l(s,R) +  q_l(s,R)  \frac{(s+1)} {\sqrt{s}}   \tan^{-1} \left(\frac{-2 %\sqrt{s}}{s - 1} \right) \right]  \!
\frac{ L^2(s + r^2) \sigma_0 / \epsilon_0}{2^{3l+2} x s^{l+1} r^{l-1}} \!
\left[  p_l(s,r) +  q_l(s,r)  \frac{(s+1)} {\sqrt{s}}   \tan^{-1} \left(\frac{-2 \sqrt{s}}{s - 1} \right) \right]  \!
\label{phil0}
\ee
\end{widetext}
where $r=R/L$, $s = (x/L)^2 - 1$, 
and $p_l, q_l$ are polynomial functions of $s$ and $r$, such that
\be
p_0(s,r) = 2\frac{s-r^2}{s+r^2} , \;\;\;\;\;\; q_0(s,r) = 1,
\ee
and for higher $l$ the polynomial is of higher order. 
The expressions for $l$ in the range $1 \le l \le 4$ are furnished in table
\ref{t.phix0}.
The value of the inverse tangent function in (\ref{phil0})
must be taken so as to obtain smooth behaviour when $s$ passes through 1.

\boxtable{t.phix0}{
%\be
%\phi(l;\,x,0,0) = \frac{ (s + R^2) \sigma_0 / \epsilon_0}{2^{3l+2} x s^{l+1} R^{l-1}} \!
%\left[  p_l(s,R) +  q_l(s,R)  \frac{(s+1)} {\sqrt{s}}   \tan^{-1} \left(\frac{-2 \sqrt{s}}{s - 1} \right) \right]  \!
%\label{phil0}
%\ee
%where $s = x^2 - 1$,
%\be
%p_0(s,R) = 2\frac{s-R^2}{s+R^2} , \;\;\;\;\;\; q_0(s,R) = 1,
%\ee
For $1 \le l \le 4$, the functions $p_l$ and $q_l$ introduced in (\protect\ref{phil0}) are given by:
\be
p_l(s,r) \!\!\!&=& \frac{4s}{3} \tilde{q}_l(s,r) - 2q_l(s,r) ,   \\
q_l(s,r) \!\!\!&=& \!\!\! 4 \sum_{m=0}^l (-1)^{m+l}  r^{2m}s^{l-m} c_{l,m} 
\left[ a s^\alpha  + b  s + (c  s + d)(1+2m)  \right]   \rule{4ex}{0pt}
\label{qlfunc}
\ee
where for $l = \{1,2,3,4\}$, the coefficients $\alpha,a,b,c,d$ are given by
\begin{eqnarray}
\alpha &=& 2   \nonumber \\  
a &=& \{0,0,0,3 \} \nonumber \\
b &=& \{0,1,8,15 \}   \nonumber\\
c &=& \{0,0,1,3 \} \nonumber\\
d &=& \{1/2,\, 1,1,5,7\}
\end{eqnarray}
and for $m = \{0,1,2,\cdots , l \} $ the coefficients $c_{l,m}$ are given by
\begin{eqnarray}
c_{1,m} &=& \{1,1 \} \nonumber\\
c_{2,m} &=& \{3,2,3 \} \nonumber\\
c_{3,m} &=& \{5,3,3,5\} \nonumber\\
c_{4,m} &=& \{ 35,20,18,20,35 \}       \label{clm}
\end{eqnarray}
and $\tilde{q}_l(s,R)$ is a function of the same form as eqn (\ref{qlfunc}), but with
\be
\alpha &=& 0   \nonumber \\  
a &=& \{3/2,0,0,0 \} \nonumber \\
b &=& \{0,0,8,15 \}   \nonumber\\
c &=& \{0,0,0,8/5 \} \nonumber\\
d &=& \{0,1,5,7\}.                   \label{abcdtilde}
\ee

%\begin{center}
%1 \\
%1,1 \\
%3,2,3 \\
%5,3,3,5 \\
%35,20,18,20,35
%\end{center}

}
{Equations (\ref{phil0})--(\ref{abcdtilde}) present the scalar potential at points on the $x$ axis inside the sphere,
for $l$ in the range $0$--$4$.
For $l>4$ the functions $p_l$ and $q_l$ continue to be polynomials of degree $l$ in $r^2$ and $s$, but the expressions for the coefficients are more complicated.}

%where
%\be
%a &=& \frac{1}{2}(l-3)(l-2)(l-1) \\
%b &=& (l-1)\left(1 + (l-2)(3 - (l-3))\right) \\
%c &=& \frac{1}{2}(l-2)(l-1) \\
%d &=& 1 + (l-1)(l-2)(2 - (l-3))
%\ee

%\frac{\sigma_0 (x^2 + R^2 - 1)}{16 R^2 x (x^2-1)^2} \left[
%2(1-3 x^2 + 2x^4 - R^2(1+2x^2)) - \frac{x^2(x^2 - 3 R^2-1)}{\sqrt{x^2-1}}
%\tan^{-1} \frac{-2 \sqrt{x^2-1}}{x^2 - 2} \right]  \\

%\frac{\sigma_0}{2 \epsilon_0}
%\frac{(s + R^2)}{8 x s^2} \left[
%2\left(2s^2 + s - R^2(1+2x^2)\right) - \frac{x^2(s - 3 R^2)}{\sqrt{s}}
%\tan^{-1} \left(\frac{-2 \sqrt{s}}{s - 1} \right) \right]  

%\end{table}

We now wish to write this potential in terms of the set of functions given in (\ref{Vnxrho}). To this
end a set of coefficients $a_{l,n}$ is defined implicitly by
\be
\phi(l;\, x,0,0) &=& \sum_{n=0}^{\infty} a_{l,n} \phi_n(x,0), \nonumber \\
\lefteqn{  L-R \le x \le L+R.}
\label{defaln}
\ee
The solution to the full problem (i.e. the potential off as well as on the axis) is then
\be
\phi(l;\, x,y,z) = \sum_{n=0}^{\infty} a_{l,n} \phi_n(x,\sqrt{y^2+z^2})
\ee
for points inside the sphere. This is the scalar potential for a multipolar sphere. When the surface charge
distribution is given by (\ref{sigseries}), the solution is
\be
\phi(x,y,z) = \sum_{l=0}^\infty s_l \, \phi(l; \, x,y,z).
\ee

The coefficients $a_{l,n}$ may be obtained from (\ref{defaln})
by any suitable method. The method
adopted was to obtain the Taylor expansions of $\phi(l;\, x,0,0)$ 
and $\phi_n(x,0)$  
about the point $x=L$, and equate coefficients of powers of $(x-L)$.
%The low-order terms of these expansions are displayed in table \ref{t.expandphi0}. 
In order to handle the infinite series in $n$, the following
strategy can be adopted. Let $c_{l,p}$ be the coefficients in the Taylor expansion of
$\phi(l;\, x,0,0)$, defined by
\be
\phi(l;\,x,0,0) = \sum_{p=0}^\infty c_{l,p} (x-L)^p,
\ee
and let $b_{n,p}$ be the coefficients in the Taylor expansion of $\phi_n$, defined
by
\be
\phi_n(x,0) = \sum_{p=0}^\infty b_{n,p} (x-L)^p.
\ee
One finds 
\be
b_{n,p} = \frac{(-1)^n}{n+1} \binomial{2n+1}{p}.
\ee
Define coefficients $a_{l,n}^{(N)}$ by
\be
\sum_{p=0}^N c_{l,p} (x-L)^p = \sum_{n=0}^N a_{l,n}^{(N)} \sum_{p=0}^N b_{n,p} (x-L)^p .
\ee
Thus the left hand side will reproduce $\phi$ when $N \rightarrow \infty$, and
for any given $N$ we can obtain  $a_{l,n}^{(N)}$ by solving the $(N+1)$ equations
\be
c_{l,p} =  \sum_{n=0}^N a_{l,n}^{(N)} b_{n,p}              \label{cMatrix}
\ee
where $0 \le p \le N$. 

\section{Self force}  \label{s.self}

The total momentum of an extended entity is defined as the sum of the momenta
of its parts, but the question arises, at what set of events is the sum to be
evaluated? This impacts on the definition of the total force on an extended body,
as discussed in the appendix. We adopt the definition given in eqns (\ref{dptot}), (\ref{dtaui}).
The electromagnetic self-force of the spherical shell is then given by
\be
\vf_{\rm self} = 
 \oint \frac{x}{L} \sigma \frac{\vE_- + \vE_+}{2}  {\rm d}S
\; =\; \vf_\sigma +
 \oint \frac{g x}{c^2} \sigma  \vE_- \dS    \label{finteg}
\ee
where $g = c^2/L$ is the proper acceleration of the centre of the sphere, 
$\vE_\pm$ are the fields owing to the sphere at its 
exterior and interior surfaces,
the integral is over the surface of the sphere, and
\be
\vf_\sigma = \oint \frac{x}{L} \frac{\sigma^2}{2 \epsilon_0} \frac{\Vr}{R}  \dS
\;=\; \frac{\hat{\rm x}}{2 \epsilon_0 R L}  \oint  \sigma^2  x (x-L) \dS
\ee
where $\Vr$ is the vector $(x-L,\,y,\,z)$. (The second equality
in (\ref{finteg}), where $\vf_\sigma$ is introduced, follows from using the Maxwell equations
to relate $\vE_+$ to $\vE_-$; see \cite{14Steane}.)

In the case of the multipolar sphere, i.e. $\sigma = \sigma_l$ for some $l$, one finds
\be
f_\sigma = \frac{2 \pi R^3 \sigma_0^2}{\epsilon_0 L} \frac{(2l^2 + 2l - 1)}{(2l-1)(2l+1)(2l+3)}.
\label{fsigma}
\ee
We postpone to section \ref{s.general} the more general charge distribution.

The second term in (\ref{finteg}) is evaluated using the series expansion described in the previous
section. One has
\be
 \oint \frac{g x}{c^2} \sigma_l  \vE_- \dS    =  \hat{\vx} \sum_{n=0}^\infty a_{l,n} f_{n,l}  \label{fsum}
\ee
where
\be
f_{n,l} = \frac{2\pi R}{L} \int_{L-R}^{L+R} x \sigma_l(\theta)  E_{x,n}\left(x, \rho(x)\right) \,{\rm d}x
\label{fsn}
\ee
where $\rho(x) = \sqrt{R^2 - (x-L)^2}$ and $E_{x,n}$ is given by eqn (\ref{Efieldn}).
This expression can be interpreted as the contribution to the self-force
owing to the order-$n$ contribution to the electric field acting on a charge distribution $\sigma_l$.
It is a central feature of the calculation that the functions $f_{n,l}$ can be obtained in closed
form. The evaluation of the integral in (\ref{fsn}) is discussed in the appendix. Some example
cases are tabulated in table \ref{t.fnl}.

As a consistency check, and for the avoidance of confusion over physical dimensions, note
that in eqn (\ref{defaln}) the function $\phi$ on the left is the complete potential, whereas the functions
$\phi_n$ on the right are simply polynomials in $x$ and $\rho$, therefore the coefficients $a_{l,n}$ are not dimensionless.
Each $a_{l,n}$ is proportional to $\sigma_0$, therefore the self-force given by (\ref{fsum}), (\ref{fsn}) 
is proportional to $\sigma_0^2$ as expected. More generally, the quantities $E_{x,n}$ have the dimensions
$L^{2n}$ and $a_{l,n}$ have the dimensions $\sigma_0 / \epsilon_0 L^{2n}$ if we adopt $L$ as
the length scale. In practice it is convenient to adopt distance units such that $L=1$.

In order to present the results, we shall introduce the following (non-standard) notation:
\be
((p))_k \equiv p (p+2)(p+4) \cdots (p+2k-2).
\ee
In terms of the Pochhammer symbol $(p)_k \equiv p(p+1) \ldots (p+k-1) = \Gamma(p+k) / \Gamma(p)$, 
this can be written
\be
((p))_k \equiv 2^k (p/2)_k.
\ee

For the case of a multipolar sphere, the final result, after including both terms in (\ref{finteg}), is
\be
f_{\rm self} &=& \frac{ 4 \pi R^3 \sigma_0^2 }{\epsilon_0 L} S_{l,l}(R/L)     \label{fresult}
\ee
where
\begin{widetext}
\begin{eqnarray}
S_{0,0} &\!=\!&
 \frac{-1}{2}
-2 \sum_{n=1}^{\infty} \frac{1}
{((t-3))_3 (t+1)} \frac{R^{t}}{L^t} \label{S0} \\
S_{1,1} &\!=\!&
\frac{-1}{18 } 
-2 \sum_{n=1}^{\infty} \frac{ (t-1) \left(t^2+t-3\right) }
{((t-5))_5 ((t+1))_2} \frac{R^{t}}{L^t} \label{S1} \\
S_{2,2} &\!=\!&
\frac{-1}{50 } 
-2 \sum_{n=1}^{\infty} \frac{t^6+t^5-3 t^4+14 t^3-217 t^2-111 t+855}
{((t-7))_7 ((t+1))_3} \frac{R^{t}}{L^t} \label{S2} \\
S_{3,3} &\!=\!&
\frac{-1}{98 } 
-2 \sum_{n=1}^{\infty} \frac{ (t-1) \mbox{poly}({\rm cf}; t) }
{((t-9))_9 ((t+1))_4} \frac{R^{t}}{L^t}, \label{S3} \\
\lefteqn{{\rm cf} =\{1, 4, 28, -8, -3164, -2372, 46884, 13590,  -137025\} } \nonumber \\
S_{4,4} &\!=\!&
\frac{-1}{162 } 
-2 \sum_{n=1}^{\infty}  \frac{\mbox{poly}({\rm cf}; t)}
{((t-11))_{11} ((t+1))_5} \frac{R^{t}}{L^t}, \label{S4}  \\
\lefteqn{ {\rm cf} = 
\{ 1, 6,116,-306,-20091,15192,686624,-921354,-5277395,} \nonumber \\
\lefteqn{ \;\;\;\;\;\;\;\; 16015662,-27687600,-53736480,159256125\} } \nonumber \\
\ldots  \nonumber
\end{eqnarray}
\end{widetext}
where $t=2n$ and the polynomials in equations (\ref{S3}), (\ref{S4}) have been indicated by listing the coefficients
of powers of $t$ in the order highest to lowest power.
$S_{0,0}$ was already calculated in \cite{Steane15}; the other results are new. 

The method of calculation involves a certain limitation on what has been proved, as opposed to what can
reasonably be conjectured. If one trusts computer algebra, then I claim to know
that the sums given in (\ref{S0})--(\ref{S4}) give the terms correctly up to the highest order that was
obtained in a symbolic calculation with the aid of the Mathematica software package. I  
conjecture that the expressions then give all the terms correctly. To prove this conjecture, it would suffice to
show that the coefficients of powers of $R$ in the expressions for $S_{l,l}$ are indeed
polynomials in $t$ of the stated order; the computer algebra was sufficient to obtain the correct 
polynomials under that condition.

%\lefteqn{\mbox{poly}(t)=t^8+4 t^7+28 t^6-8 t^5-3164 t^4-2372 t^3+46884 t^2+13590 t-137025 }  \nonumber \\
%\lefteqn{\mbox{poly}(t) =(1, 4, 28, -8, -3164, -2372, 46884, 13590,  -137025) \cdot (t^8, t^7 \ldots, 1)} \nonumber \\
% \lefteqn{\mbox{poly}(t) = t^{12}+6 t^{11}+116 t^{10}-306 t^9-20091 t^8+15192 t^7+686624 t^6 -921354 t^5 }  %\nonumber \\
% \lefteqn{ \;\;\;\; -5277395 t^4+16015662 t^3-27687600 t^2-53736480 t+159256125 }  \nonumber  \\

The lowest order terms in these series are displayed in table \ref{t.Sllseries}, and further information
is provided in the appendix.

\boxtable{t.Sllseries}
{\begin{eqnarray*}
S_{0,0} &=&
-\frac{1}{2 } + \frac{2 R^2}{9}-\frac{2 R^4}{75}-\frac{2 R^6}{735} -\frac{2 R^8}{2835} - \ldots\\
S_{1,1} &=&
-\frac{1}{18 }-\frac{2 R^2}{225}+\frac{34 R^4}{1225}-\frac{26 R^6}{3969} -\frac{46 R^8}{49005} - \ldots\\
S_{2,2} &=&
-\frac{1}{50 }-\frac{2 R^2}{2205}-\frac{6 R^4}{1225}+\frac{2042 R^6}{160083} -\frac{183914 R^8}{57972915} - \ldots\\
S_{3,3} &=&
-\frac{1}{98 }-\frac{22 R^2}{99225}-\frac{2522 R^4}{4002075}-\frac{2386 R^6}{819819}+\frac{20782 R^8}{2760615} 
- \ldots\\
S_{4,4} &=&
-\frac{1}{162 }-\frac{38 R^2}{480249}-\frac{42206 R^4}{225450225}-\frac{57802 R^6}{135270135} -\frac{10823374 R^8}{5584724145} + \ldots
\end{eqnarray*}
}
{Low-order terms in the expressions for $S_{l,l}$ as given by eqs (\ref{S0})--(\ref{S4}) at $L=1$.}

\section{Checks} \label{s.checks}

The results given above were subjected to two checks for consistency and correctness. 
First, we calculate the leading term in $S_{l,l}(R)$ by analysis, then we calculate
the whole force approximately by numerical integration. 

Nodvik showed that the lowest order contribution to the self-force, in the case of a spherically
symmetric charge distribution, is given by
\be
-\frac{1}{2}\frac{g e^2}{c^2}  \left\langle \frac{1}{| \Vr - \Vr' |} \right\rangle  \label{leading}
\ee
where $e^2 = q^2/4\pi\epsilon_0$ for a total charge $q$ and
\be
\left\langle \frac{1}{| \Vr - \Vr' |} \right\rangle = \int_{-\infty}^\infty {\rm d}^3 \Vr' \int_{-\infty}^\infty {\rm d}^3 \Vr
\frac{ f(\Vr) f(\Vr')  }{| \Vr - \Vr' |},   \label{meaninv}
\ee
where $f(\Vr)$ is the form factor describing the charge distribution. From the studies in \cite{03Ori,04Ori,14SteaneA}, 
one expects that this result also applies to a non-spherically symmetric distribution, since the effect of
departures from spherical symmetry come in at higher order in $R$. Therefore we can use (\ref{leading}) to calculate
the leading term in $S_{l,l}(R)$ for the multipolar sphere, as long as we understand the normalization of the form factor correctly.
We replace $f(\Vr) {\rm d}r$ by $\sigma(\theta)/\sigma_0$ and $q$ by $4 \pi R^2 \sigma_0$. Then
using (\ref{sig}) in (\ref{meaninv}), we find
\be
\left\langle \frac{1}{| \Vr - \Vr' |} \right\rangle = \frac{1}{(2l+1)^2 R}.
\ee
The prefactor in (\ref{fresult}) can be expressed $g (4 \pi R^2 \sigma_0)^2 / 4 \pi \epsilon_0 c^2 R$, so this
implies that $S_{l,l}(R/L)$ will be given to first approximation by
\be
S_{l,l} (R) = \frac{-1}{2(2l+1)^2}   + O(R/L).
\ee
This agrees with the first term in the expressions given in (\ref{S0})--(\ref{S4}). This constitutes our first check.

The self-force can be obtained by numerical integration as follows. From  (\ref{finteg}) we have
\begin{widetext}
\be
f_{\rm self} = f_\sigma + 2 \pi \frac{g R^2}{c^2} \int_0^\pi \!{\rm d}\theta\,
 (L + R \cos \theta) \sigma_l(\theta) E_{x,-}^{\rm shell}\left( L,R,l; \, {\vx} \right)  \sin\theta 
\label{fselfint1}
\ee
where the vector $\vx$ is given by
\be
\vx = \left(L+R\cos\theta, \; R \sin\theta, \; 0 \right) 
\ee
and $E_{x,-}^{\rm shell}$ is the $x$-component of the interior electric field of the shell, at location $\vx$. This is
given by
\be
E_{x,-}^{\rm shell}(L,R,l;\, \vx) = \lim_{h \rightarrow 0} 2\pi R^2 \! \int_0^\pi
\!{\rm d}\theta\,
\sigma_l(\theta) E^{\rm ring}(L+R_h\cos\theta, \, R_h \sin\theta; \, \vx ) \sin\theta
\ee
where $R_h = R + h$ and $E^{\rm ring}(L, R; \, \vx )$ is the $x$-component of the electric
field of a ring of unit charge, centre $(L,0,0)$ and radius $R$. This is given by
\be
E^{\rm ring}(L,R; \, \vx) = \frac{1}{2\pi} \int_0^{2\pi} 
\!{\rm d} \phi \,
E^q_x\left(L; \, \vx - (0,\, R\sin\phi,\, R \cos\phi) \right)
\ee
\end{widetext}
where
\be
E^q_x \left(L; \, \vx\right) = \frac{ 4 L^2(x_1^2 - L^2 - \rho^2)}
{ (4\pi\epsilon_0) \left[(L^2 + \vx \cdot \vx)^2 - 4 L^2 x_1^2 \right]^{3/2}}
\label{Eqx}
\ee
with $\vx = (x_1,\,y,\,z)$ and $\rho^2 = y^2 + z^2$. Here $E_x^q(L; \, \vx)$ is the $x$-component of the electric
field at $\vx$ of a unit charge undergoing hyperbolic motion along the $x$ axis, at the moment when it
comes to rest at $(L,0,0)$. \cite{60Fulton,12Steane}

\begin{figure*}
\begin{center}
\begin{tikzpicture}[scale=1] \begin{axis}[
xmin=0, xmax=1,
ymin=-3, ymax=0,
xlabel={$R$},
ylabel={$f_{\rm self} - f_\sigma$},
width=0.8 \textwidth, height=0.6 \textwidth,
]
\addplot [dotted]
coordinates {
(0.05,-0.934252)
(0.0590909,-0.790525)
(0.0681818,-0.685125)
(0.0772727,-0.604526)
(0.0863636,-0.540896)
(0.0954545,-0.489386)
(0.104545,-0.446835)
(0.113636,-0.411092)
(0.122727,-0.380645)
(0.131818,-0.354398)
(0.140909,-0.331538)
(0.15,-0.311449)
(0.159091,-0.293657)
(0.168182,-0.277788)
(0.177273,-0.263546)
(0.186364,-0.250695)
(0.195455,-0.239039)
(0.204545,-0.22842)
(0.213636,-0.218707)
(0.222727,-0.209785)
(0.231818,-0.201564)
(0.240909,-0.193963)
(0.25,-0.186916)
(0.259091,-0.180363)
(0.268182,-0.174255)
(0.277273,-0.168548)
(0.286364,-0.163204)
(0.295455,-0.158189)
(0.304545,-0.153474)
(0.313636,-0.149033)
(0.322727,-0.144842)
(0.331818,-0.140881)
(0.340909,-0.137132)
(0.35,-0.133578)
(0.359091,-0.130205)
(0.368182,-0.126998)
(0.377273,-0.123947)
(0.386364,-0.121039)
(0.395455,-0.118266)
(0.404545,-0.115618)
(0.413636,-0.113087)
(0.422727,-0.110665)
(0.431818,-0.108345)
(0.440909,-0.106122)
(0.45,-0.103989)
(0.459091,-0.101941)
(0.468182,-0.0999723)
(0.477273,-0.0980794)
(0.486364,-0.0962576)
(0.495455,-0.0945031)
(0.504545,-0.0928121)
(0.513636,-0.0911812)
(0.522727,-0.0896074)
(0.531818,-0.0880876)
(0.540909,-0.086619)
(0.55,-0.0852002)
(0.559091,-0.0838256)
(0.568182,-0.082496)
(0.577273,-0.0812083)
(0.586364,-0.0799605)
(0.595455,-0.0787506)
(0.604545,-0.0775769)
(0.613636,-0.0764376)
(0.622727,-0.0753312)
(0.631818,-0.0742562)
(0.640909,-0.0732112)
(0.65,-0.0721946)
(0.659091,-0.0712054)
(0.668182,-0.0702422)
(0.677273,-0.0693038)
(0.686364,-0.0683884)
(0.695455,-0.0674972)
(0.704545,-0.0666268)
(0.713636,-0.065777)
(0.722727,-0.0649469)
(0.731818,-0.0641356)
(0.740909,-0.0633422)
(0.75,-0.0625659)
(0.759091,-0.0618058)
(0.768182,-0.0610611)
(0.777273,-0.0603312)
(0.786364,-0.0596153)
(0.795455,-0.0589127)
(0.804545,-0.0582227)
(0.813636,-0.0575446)
(0.822727,-0.056878)
(0.831818,-0.056222)
(0.840909,-0.0555763)
(0.85,-0.0549402)
(0.859091,-0.0543132)
(0.868182,-0.0536948)
(0.877273,-0.0530845)
(0.886364,-0.052482)
(0.895455,-0.0518867)
(0.904545,-0.0512983)
(0.913636,-0.0507164)
(0.922727,-0.0501408)
(0.931818,-0.0495713)
(0.940909,-0.0490075)
(0.95,-0.0484493)
};
\addplot [ultra thin]
coordinates {
(0.05,-1.44767)
(0.0590909,-1.22496)
(0.0681818,-1.06165)
(0.0772727,-0.936767)
(0.0863636,-0.838177)
(0.0954545,-0.758367)
(0.104545,-0.69244)
(0.113636,-0.637062)
(0.122727,-0.589891)
(0.131818,-0.549227)
(0.140909,-0.513812)
(0.15,-0.482691)
(0.159091,-0.455128)
(0.168182,-0.430547)
(0.177273,-0.408489)
(0.186364,-0.388584)
(0.195455,-0.370531)
(0.204545,-0.354085)
(0.213636,-0.33904)
(0.222727,-0.325225)
(0.231818,-0.312494)
(0.240909,-0.300726)
(0.25,-0.289815)
(0.259091,-0.279671)
(0.268182,-0.270215)
(0.277273,-0.261381)
(0.286364,-0.25311)
(0.295455,-0.245348)
(0.304545,-0.238051)
(0.313636,-0.231178)
(0.322727,-0.224693)
(0.331818,-0.218564)
(0.340909,-0.212764)
(0.35,-0.207265)
(0.359091,-0.202046)
(0.368182,-0.197085)
(0.377273,-0.192364)
(0.386364,-0.187866)
(0.395455,-0.183575)
(0.404545,-0.179478)
(0.413636,-0.175561)
(0.422727,-0.171813)
(0.431818,-0.168223)
(0.440909,-0.164781)
(0.45,-0.161478)
(0.459091,-0.158306)
(0.468182,-0.155257)
(0.477273,-0.152323)
(0.486364,-0.149499)
(0.495455,-0.146778)
(0.504545,-0.144155)
(0.513636,-0.141623)
(0.522727,-0.139179)
(0.531818,-0.136817)
(0.540909,-0.134534)
(0.55,-0.132324)
(0.559091,-0.130185)
(0.568182,-0.128113)
(0.577273,-0.126104)
(0.586364,-0.124156)
(0.595455,-0.122265)
(0.604545,-0.120428)
(0.613636,-0.118644)
(0.622727,-0.116908)
(0.631818,-0.115221)
(0.640909,-0.113578)
(0.65,-0.111977)
(0.659091,-0.110418)
(0.668182,-0.108898)
(0.677273,-0.107414)
(0.686364,-0.105966)
(0.695455,-0.104552)
(0.704545,-0.103171)
(0.713636,-0.10182)
(0.722727,-0.100498)
(0.731818,-0.0992048)
(0.740909,-0.0979382)
(0.75,-0.0966973)
(0.759091,-0.0954807)
(0.768182,-0.0942876)
(0.777273,-0.0931168)
(0.786364,-0.0919674)
(0.795455,-0.0908384)
(0.804545,-0.0897289)
(0.813636,-0.0886381)
(0.822727,-0.0875652)
(0.831818,-0.0865094)
(0.840909,-0.08547)
(0.85,-0.0844464)
(0.859091,-0.083438)
(0.868182,-0.082444)
(0.877273,-0.0814641)
(0.886364,-0.0804976)
(0.895455,-0.0795443)
(0.904545,-0.0786037)
(0.913636,-0.0776754)
(0.922727,-0.0767591)
(0.931818,-0.0758547)
(0.940909,-0.074962)
(0.95,-0.0740809)
};
\addplot [dashed]
coordinates {
(0.05,-3.11155)
(0.0590909,-2.633)
(0.0681818,-2.28208)
(0.0772727,-2.01375)
(0.0863636,-1.80192)
(0.0954545,-1.63045)
(0.104545,-1.48882)
(0.113636,-1.36986)
(0.122727,-1.26853)
(0.131818,-1.18119)
(0.140909,-1.10512)
(0.15,-1.03828)
(0.159091,-0.979081)
(0.168182,-0.926289)
(0.177273,-0.878915)
(0.186364,-0.836167)
(0.195455,-0.797398)
(0.204545,-0.762077)
(0.213636,-0.729764)
(0.222727,-0.700089)
(0.231818,-0.672743)
(0.240909,-0.647461)
(0.25,-0.624017)
(0.259091,-0.602218)
(0.268182,-0.581895)
(0.277273,-0.562904)
(0.286364,-0.545116)
(0.295455,-0.528421)
(0.304545,-0.51272)
(0.313636,-0.497926)
(0.322727,-0.483962)
(0.331818,-0.47076)
(0.340909,-0.458258)
(0.35,-0.446401)
(0.359091,-0.435139)
(0.368182,-0.424429)
(0.377273,-0.41423)
(0.386364,-0.404506)
(0.395455,-0.395222)
(0.404545,-0.38635)
(0.413636,-0.377862)
(0.422727,-0.369733)
(0.431818,-0.361939)
(0.440909,-0.354459)
(0.45,-0.347274)
(0.459091,-0.340367)
(0.468182,-0.33372)
(0.477273,-0.327319)
(0.486364,-0.321149)
(0.495455,-0.315197)
(0.504545,-0.309451)
(0.513636,-0.3039)
(0.522727,-0.298533)
(0.531818,-0.293341)
(0.540909,-0.288315)
(0.55,-0.283445)
(0.559091,-0.278724)
(0.568182,-0.274145)
(0.577273,-0.269701)
(0.586364,-0.265384)
(0.595455,-0.26119)
(0.604545,-0.257112)
(0.613636,-0.253144)
(0.622727,-0.249283)
(0.631818,-0.245522)
(0.640909,-0.241858)
(0.65,-0.238285)
(0.659091,-0.234802)
(0.668182,-0.231402)
(0.677273,-0.228083)
(0.686364,-0.224842)
(0.695455,-0.221675)
(0.704545,-0.218579)
(0.713636,-0.215552)
(0.722727,-0.21259)
(0.731818,-0.209692)
(0.740909,-0.206854)
(0.75,-0.204075)
(0.759091,-0.201352)
(0.768182,-0.198684)
(0.777273,-0.196068)
(0.786364,-0.193502)
(0.795455,-0.190986)
(0.804545,-0.188516)
(0.813636,-0.186093)
(0.822727,-0.183714)
(0.831818,-0.181377)
(0.840909,-0.179083)
(0.85,-0.176829)
(0.859091,-0.174614)
(0.868182,-0.172438)
(0.877273,-0.170299)
(0.886364,-0.168197)
(0.895455,-0.166131)
(0.904545,-0.1641)
(0.913636,-0.162103)
(0.922727,-0.16014)
(0.931818,-0.15821)
(0.940909,-0.156314)
(0.95,-0.15445)
};
\addplot [ ]
coordinates {
(0.05,-13.3222)
(0.0590909,-11.2689)
(0.0681818,-9.76263)
(0.0772727,-8.61029)
(0.0863636,-7.70012)
(0.0954545,-6.96294)
(0.104545,-6.35361)
(0.113636,-5.84145)
(0.122727,-5.40488)
(0.131818,-5.02824)
(0.140909,-4.69994)
(0.15,-4.4112)
(0.159091,-4.15523)
(0.168182,-3.92672)
(0.177273,-3.72144)
(0.186364,-3.53599)
(0.195455,-3.36762)
(0.204545,-3.21403)
(0.213636,-3.07335)
(0.222727,-2.944)
(0.231818,-2.82464)
(0.240909,-2.71414)
(0.25,-2.61153)
(0.259091,-2.51599)
(0.268182,-2.4268)
(0.277273,-2.34333)
(0.286364,-2.26504)
(0.295455,-2.19145)
(0.304545,-2.12214)
(0.313636,-2.05674)
(0.322727,-1.99492)
(0.331818,-1.93638)
(0.340909,-1.88087)
(0.35,-1.82814)
(0.359091,-1.77799)
(0.368182,-1.73023)
(0.377273,-1.68468)
(0.386364,-1.64119)
(0.395455,-1.59962)
(0.404545,-1.55984)
(0.413636,-1.52172)
(0.422727,-1.48517)
(0.431818,-1.45009)
(0.440909,-1.41638)
(0.45,-1.38396)
(0.459091,-1.35276)
(0.468182,-1.32271)
(0.477273,-1.29374)
(0.486364,-1.26578)
(0.495455,-1.23879)
(0.504545,-1.21272)
(0.513636,-1.18751)
(0.522727,-1.16312)
(0.531818,-1.13952)
(0.540909,-1.11665)
(0.55,-1.09448)
(0.559091,-1.07299)
(0.568182,-1.05214)
(0.577273,-1.03189)
(0.586364,-1.01223)
(0.595455,-0.993125)
(0.604545,-0.974549)
(0.613636,-0.956482)
(0.622727,-0.9389)
(0.631818,-0.921785)
(0.640909,-0.905117)
(0.65,-0.888877)
(0.659091,-0.873049)
(0.668182,-0.857616)
(0.677273,-0.842563)
(0.686364,-0.827876)
(0.695455,-0.813541)
(0.704545,-0.799545)
(0.713636,-0.785875)
(0.722727,-0.772521)
(0.731818,-0.75947)
(0.740909,-0.746713)
(0.75,-0.734239)
(0.759091,-0.722039)
(0.768182,-0.710105)
(0.777273,-0.698426)
(0.786364,-0.686996)
(0.795455,-0.675806)
(0.804545,-0.66485)
(0.813636,-0.654119)
(0.822727,-0.643608)
(0.831818,-0.63331)
(0.840909,-0.623219)
(0.85,-0.613329)
(0.859091,-0.603635)
(0.868182,-0.594131)
(0.877273,-0.584812)
(0.886364,-0.575674)
(0.895455,-0.566712)
(0.904545,-0.557921)
(0.913636,-0.549299)
(0.922727,-0.54084)
(0.931818,-0.532541)
(0.940909,-0.524399)
(0.95,-0.51641)
};
\end{axis} \end{tikzpicture}
\begin{tikzpicture}[scale=1] \input{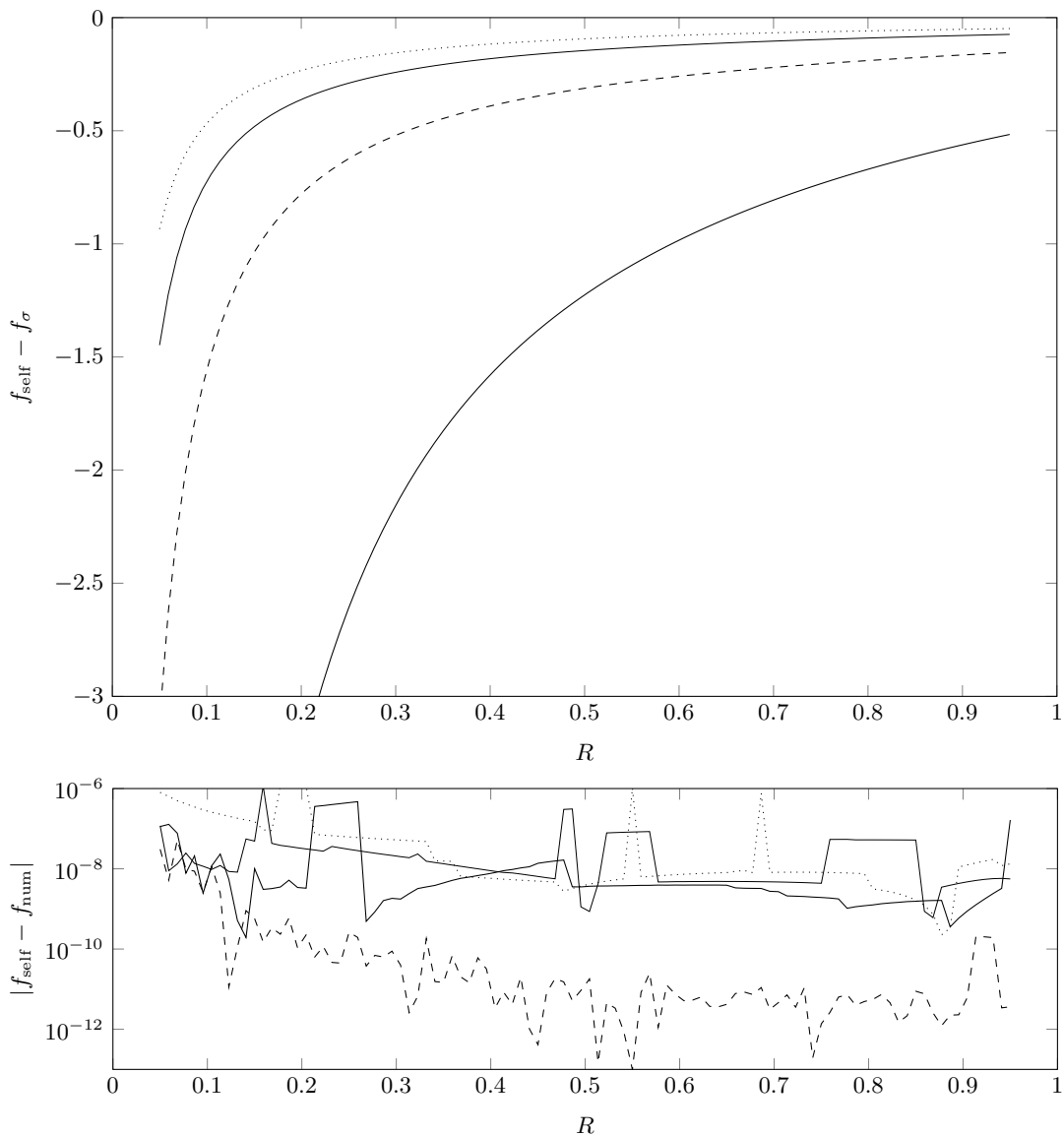} \end{tikzpicture}
\end{center}
\caption{Some example self-force results. The upper plot shows $f_{\rm self} - f_\sigma$ for
$l=0$ (thick line), $l=1$ (dashes), $l=2$ (thin line), $l=3$ (dots), as given by equations
(\protect\ref{fresult})--(\protect\ref{S3})
with $4\pi R^4 \sigma_0^2 / \epsilon_0 L= 1$.
The lower plot compares this with the results of numerical integration, by showing the magnitude of the
difference between the analytical result and a numerical integration of equations 
(\protect\ref{fselfint1})--(\protect\ref{Eqx}). The numerical estimate of the relative accuracy of the numerical
integration was $10^{-7}$. The purpose of this study was to check for possible errors or omissions in the analysis: none are found. 
The unexpectedly high precision of the numerical results at $l=1$ is fortuitous.}
\label{f.fnum}
\end{figure*}

The combination of (\ref{fselfint1})--(\ref{Eqx}) yields a triple integral and a limit extraction.
The integration was performed numerically using standard quadrature methods provided by the {\em Matlab} software
package. Care is needed owing to near-singular behaviour in the integrand for $E_{x,-}^{\rm shell}$ when $h$ is small.
The limit process was calculated by carrying out the integration at values of $h$ equal to
$(1,2,3,4,5) \times 10^{-3} R$, and fitting a quartic function to the result, so that the extrapolation to $h=0$ can
be carried out. It was difficult to get the overall relative precision below $10^{-7}$ for the integrals.
The cpu time was in the range a few minutes to one hour (depending on the
value of $R$ and $l$) to complete the calculation
for each value of $R$, when the tolerance on the relative value of the integrals was $10^{-7}$.
The integration was carried out at 100 values of $R$ in the range $0.05L$--$0.95 L$. The results matched those
of (\ref{fresult})--(\ref{S3}) to within numerical precision; see figure \ref{f.fnum}.

A further application of these numerical results is to construct a polynomial curve of best fit to the values of
$f_{\rm self}$ obtained by numerical integration, in which it is
assumed that the coefficients in the polynomial series for $S_{l,l}$
are rational numbers with denominators below some maximum
dictated by the expected numerical precision, and we assume that only non-negative even powers of $R$ are involved.
Under these assumptions one can confirm the leading terms in the series given in eqs (\ref{S0})--(\ref{S4}). However,
only a few terms can be checked this way. The requirements on numerical precision are too demanding to allow
the overall pattern in the series to be obtained.

\section{General charge distribution with axial symmetry}  \label{s.general}

We now treat a charged spherical shell whose charge distribution is
of any form having axial symmetry about the line of acceleration.
The electromagnetic self-force is given by (\ref{finteg}) where now
\be
f_{\sigma} &=& \sum_{l=0}^\infty \sum_{k=0}^\infty s_l s_k f_{\sigma}(l,k),   \\
f_{\sigma}(l,k) &=& \frac{1}{2\epsilon_0 R L}  \oint \sigma_l \sigma_{k} x(x-L) {\rm d}S   \label{fsigmagen}
\ee
and by using
\be
\vE_- = \sum_n a_n \vE_n,
\ee
where $\vE_n$ is given by (\ref{Efieldn}), (\ref{Efieldrhon}), we obtain
\be
\oint \frac{x}{L} \sigma \vE_{-} \dS = \hat{x} \sum_n \sum_l s_l a_{n} f_{n,l}.  \label{fselfgen}
\ee
Since we already know $f_{n,l}$ the problem has thus been reduced to finding
the coefficients $s_l$ and $a_n$, and performing the integral in 
(\ref{fsigmagen}). One finds
\begin{widetext}
\be
f_\sigma(l,k) =  
\frac{2\pi R^3 \sigma_0^2}{\epsilon_0 L (2l+1)(2l+3) }
\left\{ \begin{array}{ccl} 
(2l^2+2l-1) / (2l-1) && \mbox{ for } k = l,  \\
(l+1) L / R &&   \mbox{ for } k = l \pm 1,  \\
(l+1)(l+2) / (2l+5) &&   \mbox{ for } k = l \pm 2,  \\
0 &&   \mbox{ otherwise. }
\end{array}
\right.          \label{fsiglk}
\ee
\end{widetext}

Using the superposition principle (i.e. the linearity of
Maxwell's equations), we have 
$\vE_- = \sum_k s_k \vE_{k,-}$
where $\vE_{k,-}$ is the interior field owing to the $k$'th contribution to the charge.
Therefore
\be
a_n = \sum_k s_k a_{k,n}        \label{anfromsa}
\ee
where $a_{k,n}$ is the same coefficient as defined in (\ref{defaln}) and can
be obtained as discussed in section \ref{s.field}.
Using this in (\ref{fselfgen}), we have
\be
f_{\rm self} &=& \frac{ 4 \pi R^3 \sigma_0^2 }{\epsilon_0 L} 
\sum_{l=0}^\infty \sum_{k=0}^\infty s_l s_k
S_{k,l}    \label{fgenresult}
\ee
where 
\be
S_{k,l} = \left(\frac{ 4 \pi R^3 \sigma_0^2 }{\epsilon_0 L}\right)^{-1} 
\left( f_\sigma(k,l) + \sum_n a_{k,n} f_{n,l} \right)         \label{Sa}
\ee
The method has handled the non-linearity of the
problem by regarding the charge distribution $\sigma$ as a sum of distributions $\sigma_l$; 
one calculates the force on each contribution $\sigma_l$ owing to the field sourced by each 
other (or the same) contribution $\sigma_k$, and sums the results.

Using (\ref{Sa}) each $S_{k,l}$ is obtained from (\ref{fsiglk}) and (\ref{fnlu}) and the coefficients $a_{k,n}$.
To be precise, one uses not $a_{k,n}$ but $a^{(N)}_{k,n}$, obtained 
by solving (\ref{cMatrix}) by a matrix inversion, at some finite $N$, and noting that one thus
finds the result for $S_{k,l}$ up to some finite order in $(R/L)$. One then seeks to identify
the pattern in the coefficients of powers of $(R/L)$. $N$ must be chosen high enough to allow this.
On the hypothesis that the patterns thus obtained persist to all orders, one finds
\be
%S_{0,1} + S_{1,0} &=& 4 \sum_{n=1}^\infty \frac{(n-1)(2t-1)}
%{((t-5))_4 ((t-1))_2}  \frac{R^{t-1}}{L^{t-1}} \\
S_{0,1} + S_{1,0} &=&\!  \sum_{n=1}^\infty \frac{ 2t (2t + 3)}
{((t-3))_4 ((t +1))_2}  \frac{R^{t+1}}{L^{t+1}} \\
S_{0,2} + S_{2,0} &=&\!  \sum_{n=1}^\infty \frac{  -2 \left(2 t^3 - t^2 -9\right) }
{((t-5))_5 ((t+1))_2}  \frac{R^{t}}{L^{t}} \\
S_{0,3} + S_{3,0} &=&\!  \sum_{n=1}^\infty \frac{ 4 t^2 \left(t^3+4 t^2+5 t-25\right) }
{((t-5))_6 ((t+1))_3}  \frac{R^{t+1}}{L^{t+1}} \\
S_{0,4} + S_{4,0} &=&\!  \sum_{n=1}^\infty \frac{-4 t\left(t-2\right)^2  \left(t^3+3 t^2-85\right)  }
{((t-7))_7 ((t+1))_3}  \frac{R^{t}}{L^{t}}\;\;\;
\ee
\begin{widetext}
\be
S_{1,2} + S_{2,1} &\!\!=&\!\!  \sum_{n=1}^\infty \frac{ 4 t \left(t^4+5 t^3+5 t^2+t-30\right)  }
{((t-5))_6 ((t+1))_3}  \frac{R^{t+1}}{L^{t+1}} \\
S_{1,3} + S_{3,1} &\!\!=&\!\!  \sum_{n=1}^\infty \frac{-2 (t-1) \left(2 t^5+3 t^4-10 t^3-48 t^2-280 t+675\right) }
{((t-7))_7 ((t+1))_3}  \frac{R^{t}}{L^{t}} \\
S_{1,4} + S_{4,1} &\!\!=&\!\!  \sum_{n=1}^\infty \frac{ 2 t^3 \left(2 t^5+17 t^4+38 t^3-208 t^2-1264 t+2450\right) }
{((t-7))_8 ((t+1))_4}  \frac{R^{t+1}}{L^{t+1}}
\ee
\be
S_{2,3} + S_{3,2} &=&  \sum_{n=1}^\infty \frac{ 2 t \mbox{ poly}({\rm cf}; t) }
{((t-7))_8 ((t+1))_4}  \frac{R^{t+1}}{L^{t+1}} , \\
%\lefteqn{ {\rm cf} = \{ 2, 19 , 82, 181, -1526, -5483, 7094, 8505 \} }  ; \nonumber\\
{\rm cf} &=& \{ 2, 19 , 82, 181, -1526, -5483, 7094, 8505 \}   ; \nonumber\\
S_{2,4} + S_{4,2} &=&  \sum_{n=1}^\infty \frac{ -2  \mbox{ poly}({\rm cf}; t)}
{((t-9))_9 ((t+1))_4}  \frac{R^{t}}{L^{t}} , \\
%\lefteqn{ {\rm cf} = \{ 2 ,5 ,18 ,-261 ,-4326 ,10866 ,46408 ,-129518, 163800, -198450 \} } \\
{\rm cf} &\!=&\!\!\{ 2 ,5 ,18 ,-261 ,-4326 ,10866 ,46408 ,-129518, 163800, -198450 \}  \nonumber
\ee
\be
S_{3,4} + S_{4,3} &=&  \sum_{n=1}^\infty \frac{ 4t  \mbox{ poly}({\rm cf}; t)}
{((t-9))_{10} ((t+1))_5}  \frac{R^{t+1}}{L^{t+1}} , \\
{\rm cf} &\!=&\!\!\{
1,15,147,627,-6786,-45750, 72338,  483588,-163980,-191700,-1701000\}  \nonumber
\ee
\end{widetext}

Expressions for higher $k,l$ can be obtained as needed. The only integral to be performed is (\ref{phix0}). 
This integral is straightforward, if laborious, at any given $l$, but we have not found an explicit general form for
the outcome. The rest of the algebra required to obtain $S_{l,k}$
is laborious, but each step is simple and can be automated. Since this calculation only
needs to be done once for each $l,k$, it supplies, in principle, the means to treat a general $\sigma(\theta)$.
One thus reduces the whole problem of finding the self-force of the charged shell to that of
obtaining the coefficients $s_l$, which is to say, the weights of the various multipole moments of the
charge distribution.

\section{Applications} \label{s.application}

\subsection{The dipole}  \label{s.dipole}

In the history of the subject, both the monopole and the dipole have yielded important insight into the physics of self-force. 
In particular, a physical object consisting of two small oppositely 
charged spheres separated by a short rod was discussed. \cite{83Griffiths,86Cornish,86Griffiths,03Ori,06Pinto}
This proved important
because it led to two paradoxes. One paradox is owing to the fact that, in such a case, 
when the whole system is accelerating in the direction orthogonal to the line between the spheres, 
the force exerted by each sphere on the other is directed somewhat in the {\em forward} direction (i.e. the 
same direction as the acceleration), and this
led to the suggestion that the system can accelerate even in the absence of any externally applied force. 
\cite{86Cornish,86Griffiths}
If such self-acceleration were possible then it would violate energy and momentum conservation. The second paradox is
owing to the fact that the electromagnetic self-force, and consequently the contribution to inertia, 
can depend on the orientation of the dipole relative to its acceleration at lowest order, if it is calculated
a certain way, whereas the field energy does not,
which suggests that the momentum and energy of the dipole plus its field cannot respect the
principle of relativity.  \cite{83Griffiths,06Pinto} 

The first paradox is resolved by noting that each sphere also exerts a self-force on itself, and 
this self-force is larger than that owing to the other sphere, and in the opposite direction. 
This was shown by analysis in the limit where the spheres
are small compared to their separation, and by numerical integration in some other cases. \cite{14SteaneA} 
This simple resolution had previously been ignored owing to the practice of absorbing the 
lowest-order term in the self-force into the definition of the body's mass. Such a practice is 
not in itself inappropriate but it can result in one failing
to notice when an unphysical assumption has been made about the relation of the mass to the size of the body \cite{14SteaneB}.

The second paradox is resolved by remembering to include the effect of 
stress in the material of the rod: this contributes to the inertia.  \cite{14SteaneA} (This is the 
effect that can also be described through the concept of `hidden momentum'.) 
A misunderstanding very closely related to this one has been treated by Barnett and resolved in the same way
\cite{Barnett13}. A related issue is the force on an inertially moving magnetic dipole; this too
can be calculated and hence interpreted in more than one way. The contribution of hidden momentum must 
be included either explicitly or implicitly, for example by using the correct Lagrangian; see \cite{Hnizdo12}
and works cited therein.

In view of the fact that the dipole has served this instructive role, and in view of the fact that it is a
simple case that one would naturally like to understand, there is interest in calculating---exactly if possible---the self-force 
of a dipole-like distribution of charge. Equation (\ref{S1}) gives the result of such a calculation. It only treats one orientation of the dipole relative to its direction of acceleration, but it
furnishes an exact closed expression. This has not previously been achieved.

\subsection{Minimum self-force for a given charge}  \label{s.min}

With closed expressions in our possession, we are well-placed to explore 
further questions about the physics
of self-force. For example, for a spherical shell carrying a given net charge, one may ask: how does the self-force
depend on the way the charge is distributed? If the charge is concentrated into a small
region of the shell, the self-force will increase. Conversely, if the charge is spread out then there will exist
a distribution which minimises the self-force. One might suppose that a uniform
distribution would minimize the electromagnetic self-force. 
This is not always so, as we now show.

Consider the charge distribution $\sigma = \sigma_0(1 + \mu \cos \theta)$
where $\sigma_0$ is fixed and $\mu$ is allowed to vary. The total charge is then $q = 4 \pi R^2 \sigma_0$
and the self-force is proportional to $S_{0,0} + \mu^2 S_{1,1} + \mu( S_{0,1} + S_{1,0} )$.
The limit of small acceleration is the limit
 $L \rightarrow \infty$. In this limit the term proportional to $\mu$ vanishes in comparison with the
others, and therefore in this case the magnitude of the self-force is smallest for the uniform distribution
($\mu = 0$). It follows that the observed mass of the charged
spherical shell is smallest for a uniform distribution of charge, if we define the observed mass as the
ratio of applied force to acceleration in the limit of small acceleration, and we assume
a possible contribution from sheer stress in the material of the shell does not overturn this conclusion.

At larger acceleration, on the other hand, the term
linear in $\mu$ is non-negligble, and consequently the self-force reaches its smallest magnitude
for some non-zero value of $\mu$ which depends on $R/L$. The minimum is at
\be
\mu = - \frac{ S_{0,1} + S_{1,0} }{2 S_{1,1}}.
\ee
For given $R$, the largest acceleration possible yields $L=R$, because at higher acceleration the body
cannot remain rigid (it would extend over the horizon in Rindler space). In this case (c.f. the appendix)
\be
\mu = -\frac{6 \left(\pi ^2-16\right)}{3 \pi ^2-64} \simeq -1.07.  \label{mu}
\ee
This represents a charge distribution with the charge slewed towards low $x$, which is the region where
the acceleration is highest---a counter-intuitive result. It suggests, for example, that
the weight of a charged object centred at a given location
near the horizon of a black hole is smaller when the charge 
is distributed nearer to the horizon, which seems to contradict the general observation that the electromagnetic
self-force, and hence the weight, increases when the gravitational field strength increases, for a  charged
body at rest in a gravitational field. 

There is no contradiction in fact. 
The intuitive sense of surprise results from the ambiguity produced when any 
non-invariant quantity is discussed without
noting its dependence on the choice of reference frame (inertial or otherwise). In the present context,
the self-force is not a property of the body alone, but is a statement about momentum changes
between chosen hypersurfaces, and consequently references to `the weight' of a body are 
ambiguous until the reference worldline and hypersurfaces have been specified. 
The counter-intuitive result given by (\ref{mu}) is owing to the scaling factor $x/L$ in (\ref{finteg}),
which comes from (\ref{dptot}), (\ref{dtaui}). 
For a given increment of proper time at the centre of the sphere, the increment
in proper time at low $x$ is smaller than at high $x$, for the set of hypersurfaces used to calculate the
force, and consequently the region at small $x$ gets a lower weighting 
in the calculation of the total force. This is why the total force, as we have defined it, 
goes down when the charge is displaced somewhat towards low $x$. However, this does not 
necessarily
imply that the body then becomes easier to support, because the calculation of the force provided
by whatever system is used to support the body would be subject to the same scaling.

More generally, as one approaches a horizon, contributions to the sum in (\ref{dtaui})
that are nearer the horizon have a smaller value of ${\rm d}\tau_i/{\rm d}\tau_c$ and thus have a
lower weighting in the sum, but this statement applies equally to all (non-gravitational) forces acting on the body.
Such considerations bear on the study of forces and energy movements when an object is lowered
gradually into a black hole. \cite{06Rindler,12Steane}

%The paradoxical prediction that we would like to understand is the
%prediction that the total force is smaller when the charge is predominantly located in the place
%where the effects of acceleration are larger.

\section{Conclusion}

This paper solves the problem of electromagnetic self-force for the axially symmetric charged
spherical shell. The method can in principle be generalised to an arbitrary charge distribution. 
An axially symmetric charge distributed over a three-dimensional region (i.e. not confined to
a shell) can be broken down into a set of concentric shells in an obvious manner. To find the self-force,
one would then
require the field exterior to, as well as interior to, each shell. These can be found by a modest
extension of the methods of this paper. Avoiding the restriction to axial symmetry is
more difficult. It would require a more general set of basis functions, and the integrals required
in order to express the scalar potential in terms of these functions
would be much more difficult.

The paper has treated rigid motion at constant proper acceleration,
without specifying how the forces giving rise to that motion might arise. This is somewhat artificial, since
it would require a very particular set of stresses in the surface of the shell, combined with whatever
is the externally applied force, to ensure that each part of the shell gets the acceleration that has been assumed. 
If would be useful to determine precisely what those stresses are, for an example case such as motion
in a constant uniform applied electric field.

\section{Appendix A}

\subsection{Defining total force on an extended body}

Let $\chi$ be a spacelike hypersurface,
and model an extended body as a set of small parts $i$. The total 4-momentum of the body is defined to be
\be
p^\mu_{\rm tot}(\tau_c,\chi) = \sum_{i}  p_i^\mu \left(\tau_{i,\chi}\right) . \label{sump}
\ee
where $\tau_{i,\chi}$ is the proper time on the $i$'th worldline when that worldline intersects $\chi$,
and $\tau_c$ is the proper time on some reference worldline (e.g. the worldline of the centroid). When
the body is isolated, the conservation of
energy and momentum has the result that $p^\mu_{\rm tot}(\tau_c,\chi)$
is independent of $\chi$. More generally this is not guaranteed and therefore
one must specify $\chi$ when referring to the total momentum of an extended system.
Typically, one picks a spacelike hyperplane (so that the events $\{i\}_\chi$ are
simultaneous in some frame).

The total 4-force is given by \cite{64Nodvik,82Pearle,03Ori,15Harte,74Dixon,14SteaneA,Steane15}. 
\be
\dby{\ffp^\mu_{\rm tot}}{\tau_c} &=& 
 \sum_i \dby{\ffp^\mu_i}{\tau_i}  \dby{\tau_i}{\tau_c}   \label{dptot}
\ee
where each ${\rm d}\tau_i$ is the proper time elapsed on the $i$'th
worldline between the intersections of that worldine with $\chi$ and $\chi+{\rm d}\chi$,
and  the quantities ${\rm d}\ffp^\mu_i/\dtau_c$
and $\dtau_i / \dtau_c$ are evaluated on the hyperplane $\chi$.

For an object undergoing rigid motion there is a natural choice of $\chi$,
namely the hypersurface orthogonal to all the worldlines at $\tau_c$. This is
the choice we shall make here.  For the
hyperplane $\chi + {\rm d}\chi$ one may choose the plane parallel to
$\chi$ and intersecting the reference worldline at $\tau_c + \dtau_c$, or
one may choose the plane orthogonal to the worldlines (among other possible
choices). In the first case, $\dtau_i / \dtau_c = 1$, and in the second
\be
\dby{\tau_i}{\tau_c} = \frac{x}{x_c}  \label{dtaui}
\ee
%$\dtau_i /\dtau_c = x/x_c$
for the motion under consideration here (rigid motion at constant proper acceleration). A
suitable reference point is the centre of the sphere, giving $x_c =L = c^2/g$ where
$g$ is the proper acceleration of the centre of the sphere.

The treatments given in \cite{64Nodvik,03Ori,82Pearle} all adopt the
second choice (i.e. eqn (\ref{dtaui})) for the purpose of defining
and calculating
self-force for a charge distribution undergoing rigid motion. Ori showed
that, with this choice of $\chi + {\rm d}\chi$, the self-force 
has the following desirable feature: for a pair of charges at opposite ends of
a straight rod of fixed proper length and centred at $x_c$, the contribution
to the total self-force owing to the field of each charge at the other is 
independent of the orientation of the rod, to lowest order in the length of
the rod. If one defines the self-force through some other choice of
hypersurface, this feature will not in general hold, and then in order to make
sense of the dependence on orientation one must take into account the internal
stress in the body supporting the charges \cite{14SteaneA}. In this 
connection, Steane \cite{Steane15}
showed a further desirable property of (\ref{dtaui}): with this choice, 
the contribution to
the self-force owing to internal pressure is independent of the orientation
of the body, if the equation of state of the interior of the body is that of an
ideal fluid (that is, one exhibiting tension or pressure but not sheer stress).

\subsection{Evaluating $f_{n,l}$}

We present the calculation of $f_{n,l}$ (eqn (\ref{fsn})). 

Using
\be
E_{x,0} &=& - 2, \\
E_{x,1} &=& -2(\rho^2 - x^2), \\
E_{x,2} &=& -2(\rho^4 - 4 x^2 \rho^2 + x^4),
\ee
it is straightforward to obtain the values of $f_{n,l}$ for the lowest values of $n$ and $l$ by performing the integral in eqn (\ref{fsn}). They are
\be
f_{0,0} &=& (4\pi R^2 \sigma_0) 2 (-1) , \\
f_{0,1} &=& (4\pi R^2 \sigma_0) 2  (-1 / 3 )R/L ,\\
f_{1,0} &=& (4\pi R^2 \sigma_0) 2 (L^2 + R^2/3) ,\\
f_{1,1} &=& (4\pi R^2 \sigma_0) 2  (L^2 + R^2/15) R/L .
\ee
When the factor $(4 \pi R^2 \sigma_0)$ is thus taken to the front of the expression, the
rest of the expression gives the dependence on $R$ and $L$ in the case of a sphere carrying
a fixed amount of charge in each fraction of its surface as $R$ varies.

More generally, by substituting (\ref{Efieldn}) into (\ref{fsn}), we obtain
\be
f_{n,l} = -8\pi R^2  \sigma_0 L^{2n} \sum_{k=0}^n (-1)^{n+k} \binomial{n}{k}^{\!2} \! \left( \frac{R}{L} \right)^{\! 2k} \!
 J_{n,k,l}      \label{fnlu}
\ee
where
\be
J_{n,k,l} = \frac{1}{2} \int_{-1}^1 \left( \!\frac{R u}{L} + 1 \! \right)^{\! 2(n-k)+1} \! (1-u^2)^k P_l(u) \,{\rm d}u . \; \label{Jnkldef}
\ee 
The values of $f_{n,l}$ for $n$ in the range 0--4 and $l$ in the range 0--5 are shown in table \ref{t.fnl}. The following general observations may be made. 
One finds $f_{n,l} = 0$ for $n < (l-1)/2$. For even(odd) $l$, only even(odd) powers
of $(Ru)$ in the integrand contribute to the answer.  When $L=1$, 
$f_{n,l}$ is a polynomial in $R$, with lowest order term of order 
$(4\pi R^2\sigma_0) R^{l}$. After dividing $f_{n,l}$ by this, the expression that remains is a polynomial in $R^2$
of order $n - \lfloor l/2 \rfloor$. 

\begin{table*}
\[
\begin{array}{l|ccccc}
l & n=0& 1 & 2 & 3 & 4 \\
\hline
0\rule{0pt}{2.5ex}& -1 & \frac{R^2}{3}+1 & \frac{R^4}{15}-\frac{2 R^2}{3}-1 & \frac{R^6}{35}-\frac{R^4}{5}+R^2+1 & \frac{R^8}{63}-\frac{4
   R^6}{35}+\frac{2 R^4}{5}-\frac{4 R^2}{3}-1 \\
1\rule{0pt}{2.5ex}& -\frac{R}{3} & \frac{R^3}{15}+R & \frac{R^5}{105}-\frac{2 R^3}{5}-\frac{5 R}{3} & \frac{R^7}{315}-\frac{3
   R^5}{35}+R^3+\frac{7 R}{3} & \frac{R^9}{693}-\frac{4 R^7}{105}+\frac{2 R^5}{7}-\frac{28 R^3}{15}-3 R \\
2\rule{0pt}{2.5ex}& 0 & \frac{8 R^2}{15} & -\frac{4 R^4}{21}-\frac{28 R^2}{15} & -\frac{4 R^6}{105}+\frac{32 R^4}{35}+4 R^2 & -\frac{8
   R^8}{495}+\frac{8 R^6}{35}-\frac{88 R^4}{35}-\frac{104 R^2}{15} \\
3\rule{0pt}{2.5ex}& 0 & \frac{4 R^3}{35} & -\frac{4 R^5}{105}-\frac{44 R^3}{35} & -\frac{8 R^7}{1155}+\frac{8 R^5}{15}+\frac{32 R^3}{7} &
   -\frac{8 R^9}{3003}+\frac{136 R^7}{1155}-\frac{152 R^5}{63}-\frac{56 R^3}{5} \\
4\rule{0pt}{2.5ex}& 0 & 0 & -\frac{16 R^4}{35} & \frac{208 R^6}{1155}+\frac{1072 R^4}{315} & \frac{16 R^8}{429}-\frac{5536
   R^6}{3465}-\frac{1328 R^4}{105} \\
5\rule{0pt}{2.5ex}& 0 & 0 & -\frac{16 R^5}{231} & \frac{80 R^7}{3003}+\frac{368 R^5}{231} & \frac{16 R^9}{3003}-\frac{160
   R^7}{231}-\frac{6928 R^5}{693} \\
\end{array}
\]
\caption{The value of $f_{n,l} /(8 \pi R^2 \sigma_0)$ when $L=1$, for $n = 0,1, \cdots 4$ and $l = 0,1, \cdots 5$.}
\label{t.fnl}
\end{table*}

By expanding both brackets in the integrand in (\ref{Jnkldef}) using the binomial theorem, we have
\be
J_{n,k,l} =  \sum_{p=0}^{p_{\rm max}} \sum_{q=0}^k (-1)^q 
 \binomial{p_{\rm max}}{p} \binomial{k}{q}  \left( \frac{R}{L} \right)^p I_{p + 2 q, \, l} \;\;
\ee
where $p_{\rm max} = 2(n-k)+1$ and
\be
I_{p,l} &=& \frac{1}{2} \int_{-1}^1 u^p P_l(u) {\rm d}u .
\ee
This integral is presented in Gradsteyn and Ryzhik \cite{GRyz}, eq 7.126(1), EH I 171(23) (p.771), which states that,
for ${\rm Re} \, \sigma > -1$,
\be
\int_0^1 P_\nu(x) x^\sigma {\rm d}x = \frac{ \sqrt{\pi} 2^{-\sigma-1} \Gamma(1+\sigma) }
{ \Gamma(1 + (\sigma-\nu)/2) \Gamma( (\sigma+\nu+3)/2 ) } .
\ee
I have confirmed this for a large range of values of $\nu$ and $\sigma$. The result for $I_{p,l}$ is
\be
I_{p,l} = 
\left\{ \begin{array}{ll}
0, & p < l  \\
0, & p + l \mbox{ is odd } \\
\displaystyle \frac{  (-1)^{(p - l +2)/2}  \left(p - l - 3\right)!! \, (1)_{p}  }
{ 2^{(3 p - l)/2}  ((p - l)/2)!   \left(\frac{3+l-p}{2} \right)_p  }
& \mbox{otherwise} 
\end{array}  \right.
\ee
where the two subscripted brackets are Pochhammer symbols.

%Sum[ Sum[ (-1)^q Bin[k, q] Bin[2 (n - k) + 1, p] R^p Ipl[p + 2 q,     l], {q, 0, k}], {p, 0, 2 (n - k) + 1} ]
% -2 Sum[ Bin[n, n - k]^2 (-1)^(n - k) R^(2 k) Jnkl[n, k, l], {k, 0, n} ]

%Ipl[p_, l_] := If[ p < l, 0,
%  If[ Mod[p + l, 2] == 1, 0,
%  (-1)^((p - l)/2 + 1)  2^-(l + 3 (p - l)/2) Pochhammer[1, 
%    l + (p - l)] ((p - l) - 3)!!/(((p - l)/2)!   Pochhammer[ 
%      3/2 - (p - l)/2, l + (p - l)]) ] ]

\subsection{Further information}  

The value of the electric field at the centre of the sphere, obtained from (\ref{phix0}), is 
\be
\lefteqn{E_x(L,0,0) = \frac{ (-1)^l  \sigma_0 (2l-3)!! }{2^l \epsilon_0 (2l+3)!!} \left( \frac{R}{L} \right)^{l-1} 
\left[ -\frac{2 l(2l+3) }{2l-3}    \rule{0ex}{3ex}  \right. } \nonumber \\
%&&\!\!\!\!\!  \left. - 2(l^2+l-1) \left( \frac{R}{L} \right)^{\! 2}
%+ \frac{(l+1)(l+2)^2 (2l-1) }{2 (2l+5)}\left( \frac{ R}{L} \right)^{\! 4} \right] .\;\;\;\;
&&\!\!\!\!\!  \left. - 2(l^2+l-1) \frac{R^2}{L^2} 
+ \frac{(l+1)(l+2)^2 (2l-1) }{2 (2l+5)} \frac{R^4}{L^4} \right] .\;\;\;\;
\ee
For the $l=0,1,2$ the expression evaluates to
\be
-\frac{2 r}{3} \!+\! \frac{2 r^3}{15}, \; -\frac{1}{3} \!+\! \frac{r^2}{15} -\! \frac{3 r^4}{70}, \;
-\! \frac{r}{15} - \frac{r^3}{42} \!+\! \frac{2 r^5}{105}, \; \ldots\;\;
\ee
after omitting a factor  $\sigma_0 / \epsilon_0$ and using $r = R/L$.
The limit where the acceleration goes to zero is the limit $L \rightarrow \infty$
and therefore $r \rightarrow 0$ in these expressions.
The dipole case ($l=1$) then gives $E = (-1/3) \sigma_0 / \epsilon_0$ which is the familiar electric field inside a spherical shell carrying a dipolar
distribution of surface charge. For higher $l$ there is a value of $R/L$ where $E_x(L,0,0)$ passes through zero.

%\be
%S_{0,0} &=& 
%-\frac{r^3 \Phi \left(r^2,2,\frac{3}{2}\right)-\left(r^2-5\right) r+\left(r^4-4 r^2+3\right) \tanh ^{-1}(r)}{16 r}  \\
%S_{1,1} &=&
%-\frac{2 r^5 \left(\Phi \left(r^2,2,\frac{5}{2}\right)-2 \Phi \left(r^2,2,\frac{3}{2}\right)\right)-3 (r-1) (r+1)
%   \left(r^4+1\right) r+\left(3 r^8-4 r^6+4 r^2-3\right) \tanh ^{-1}(r)}{256 r^3}-\frac{1}{18}
%\ee

In the main text, the results for $S_{k,l}$ are given as infinite series. These series can in principle be summed. One finds, for example,
\begin{widetext}
\be
S_{0,0} &=& \frac{1}{16} \left[
(r^2-5) - r^2 \Phi_2 \left(r^2,\mbox{$\frac{3}{2}$}\right) 
-\left(r^4 - 4 r^2 +3 \right) \frac{\tanh ^{-1}(r)}{r} \right]  \\
S_{1,1} &=& -\frac{1}{18} + \frac{1}{256r^2} \left[  \rule{0pt}{3ex}
3 (r^2 - 1)   \left(r^4+1\right) 
-2 r^4 \left(\Phi_2 \left(r^2,\mbox{$\frac{5}{2}$}\right)-2 \Phi_2 \left(r^2,\mbox{$\frac{3}{2}$}\right)\right) \right. 
% \nonumber\\&& 
\left. -\left(3 r^8-4 r^6+4 r^2-3\right) \frac{\tanh ^{-1}(r)}{r}  \right] \;\;
\ee
\be
S_{0,1}+S_{1,0} = 
%\rule{0.75\textwidth}{0pt} \nonumber\\
\frac{
-\left(3 r^4+5\right) r
- \left(r^2-3\right) r^3 \Phi_2 \left(r^2,\frac{3}{2}\right) 
 +\left(3 r^6 - 5 r^4 - 3 r^2 + 5\right)   \tanh ^{-1}(r) }{64 r^2}
\ee
where $r = R/L$ and $\Phi$ is the Lerch transcendent
\be
\Phi_s(z,a) = \sum_{n=0}^\infty \frac{z^n}{(n+a)^s} .
\ee
At $r=1$ these expressions give 
$S_{0,0} = -\pi^2/32,\; S_{1,1} = (3 \pi^2 - 64)/768,\;S_{0,1}+S_{1,0} = (\pi^2-16)/64$.
\end{widetext}

%\bibliographystyle{}
%\bibliographystyle{plain}
%\bibliography{selfforcerefs}

% when using arXiv use the 2nd version
\ifodd\localfiles
    \bibliographystyle{unsrt}

\else
    \bibliographystyle{unsrt}
    \bibliography{../selfforcerefs}
\fi

\end{document}